\documentclass[twocolumn]{aastex631}

\graphicspath{{./}{figures/}}

\begin{document}

\title{A Framework to Calibrate a Semi-analytic Model of the First Stars and Galaxies to the Renaissance Simulations}


\author[0000-0002-1034-7986]{Ryan Hazlett}
\affiliation{Ritter Astrophysical Research Center, Department of Physics and Astronomy, University of Toledo, 2801 W. Bancroft Street, Toledo, OH 43606, USA}
\email{ryan.hazlett@rockets.utoledo.edu}

\author[0000-0002-9789-6653]{Mihir Kulkarni}
\affiliation{Ritter Astrophysical Research Center, Department of Physics and Astronomy, University of Toledo, 2801 W. Bancroft Street, Toledo, OH 43606, USA}
\email{mihir.kulkarni@utoledo.edu}

\author[0000-0002-8365-0337]{Eli Visbal}
\affiliation{Ritter Astrophysical Research Center, Department of Physics and Astronomy, University of Toledo, 2801 W. Bancroft Street, Toledo, OH 43606, USA}
\email{elijah.visbal@utoledo.edu}

\author[0000-0003-1173-8847]{John H. Wise}
\affiliation{Center for Relativistic Astrophysics, Georgia Institute of Technology, 837 State Street, Atlanta, GA 30332, USA}
\email{jwise@physics.gatech.edu}



\begin{abstract}

We present a method that calibrates a semi-analytic model to the \emph{Renaissance Simulations}, a suite of cosmological hydrodynamical simulations with high-redshift galaxy formation. This approach combines the strengths of semi-analytic techniques and hydrodynamical simulations, enabling the extension to larger volumes and lower redshifts inaccessible to simulations due to computational expense. Using a sample of \emph{Renaissance} star formation histories (SFHs) from an average density region of the Universe, we construct a four parameter prescription for metal-enriched star formation characterized by an initial bursty stage followed by a steady stage where stars are formed at constant efficiencies. Our model also includes a treatment of Population III  star formation where a minimum halo mass and log-normal distribution of stellar mass are adopted to match the numerical simulations. Star formation is generally well reproduced for halos with masses $\lesssim$$10^{9} M_{\mathrm{\odot}}$. Between $11<z<25$ our model produces metal-enriched star formation rate densities (SFRDs) that typically agree with \emph{Renaissance} within a factor of $\sim$2 for the average density region. Additionally, the total metal-enriched stellar mass only differs from \emph{Renaissance} by about $10\%$ at $z \sim 11$. For regions that are either more overdense or rarefied not included in the calibration, we produce metal-enriched SFRDs that agree with \emph{Renaissance} within a factor of $\sim$2 at high-$z$, but eventually differ by higher factors for later times. This is likely due to environmental dependencies not included in the model. Our star formation prescriptions can easily be adopted in other analytic or semi-analytic works to match our calibration to \emph{Renaissance}.

\end{abstract}



\section{Introduction} \label{sec:intro}
Understanding the first stars and galaxies is one of the major goals of modern astronomy. In the context of the $\Lambda$-Cold Dark Matter ($\Lambda$CDM) model of cosmology, simulations and analytic calculations predict that the first Population III (Pop III) stars formed from primordial gas within ${\sim}10^8$ years after the Big Bang via molecular hydrogen cooling in ${\sim}10^{5-6}~M_\odot$ dark matter (DM) ``minihalos'' \citep{1996haiman_2,1997tegmark,2002abel,2003yoshida,2007oshea,2007wise,2015greif,2021kulkarni,2021schauer,2023nebrin,2023hegde}. Due to less efficient gas cooling during star formation, Pop III stars are expected to have higher masses than Pop II/I, perhaps with typical values of ${\sim}10-1000~M_\odot$ \citep{2011greif_2,2011hosokawa,2014hirano,2014susa,2018ishigaki}, but the precise initial mass function (IMF) remains uncertain.

We note that because the very first stars are predicted to form in small numbers at very high redshifts ($z{\gtrsim} 20$), they are unlikely to be directly observed even with the 
\emph{James Webb Space Telescope} ({\emph{JWST}}). However, most models predict that Pop III star formation should continue well into or beyond the epoch of reionization \citep{2022riaz,2023borrow} where they might be detected either as purely Pop III galaxies or in mixtures of Pop III and metal-enriched star formation (e.g., in the case of inefficient metal-mixing). In fact, there are already some exciting but tentative/unconfirmed Pop III signatures from \emph{JWST} \citep{2022wang,2023maiolino}. Other promising observational probes of Pop III stars include He \textsc{ii} line intensity mapping \citep{2015visbal,2022parsons}, pair-instability supernovae detection \citep{2013whalen,2018hartwig}, 21cm cosmology \citep{2012visbal,2013fialkov,2017cohen,2018bowman}, and stellar archaeology \citep{2015frebel}. 

Due to the hierarchical character of large-scale structure formation in $\Lambda$CDM, the minihalos hosting Pop III stars serve as building blocks for the first metal-enriched galaxies. Thus, the properties of high-redshift low-mass galaxies observable with \emph{JWST} are expected to depend on the details of Pop III star formation. For example, \cite{2021abe} found that a top-heavy Pop III initial mass function (IMF), which results in more highly energetic pair-instability supernovae (PISNe), results in suppressed star formation in galaxies hosted by ${\sim}10^9~M_\odot$ halos. 
Understanding the first metal-enriched galaxies is also interesting because they may drive the process of cosmic reionization, they represent the earliest stages of more evolved galaxies, and the existence of low-mass dark matter host halos, puts constraints on alternative dark matter models, such as ``fuzzy'' or warm dark matter \citep{2007Gao,2019Mocz,2020Mocz,2022kulkarni}, that suppress small-scale structure.
Using \emph{JWST} we are able to study these galaxies individually and determine their global properties through quantities such as the UV luminosity function (LFs) \citep{2023bouwens,2023atek}. 

Theoretical models are required to interpret and guide the observations of the first stars and galaxies described above. A primary tool which has been used is hydrodynamical cosmological simulations. State-of-the-art simulations can include a variety of physical processes such as gas chemistry/cooling, radiative transfer of UV photons, and feedback from supernovae explosion. These effects can be modeled with high fidelity, but have the drawback of being numerically expensive. For instance, the \emph{Renaissance Simulations}, described in detail below, used around 10-million CPU hours on the NCSA Blue Waters supercomputers to run a box size of 135 Mpc$^{3}$ to a redshift of $z \sim 15$ \citep{2014xu,2014chen}.

This presents a serious challenge because in order to make predictions for the observations described above one needs much larger volumes than can be modeled with hydrodynamical simulations. For example, in order to understand the impact of Pop III stars on low mass galaxies, one needs a large-statistical sample of galaxies forming in ${\sim}10^9~M_\odot$ DM halos, but also needs to resolve the low-mass halos that host Pop III star formation. Despite its computational expense, the \emph{Renaissance} volume that can be simulated only contains a few thousand of these objects, much less than required for robust statistical analysis.

A different technique is to use computationally efficient semi-analytic models which combine analytic prescriptions with dark matter halo merger trees generated from $N$-body simulations or Monte Carlo methods. Semi-analytic simulations have been used extensively to study the first stars and galaxies \citep[e.g.,][]{2009trenti,2013crosby,2018visbal,2018mebane,2019yung,2020liu}. These simulations are able to rapidly scan the relevant parameter space (e.g., the Pop III IMF and critical metallicity for Pop III star formation). However, the main drawback of this type of modelling is that a large number of free parameters are often employed and it can be unclear to what extent analytic assumptions are justified.

In this paper, we explore using the  \emph{Renaissance} simulations to perform a detailed calibration of a semi-analytic model of the first stars \citep{2020visbal}. The goal of the exercise is to obtain a model that is both computationally efficient and well-matched to detailed numerical simulations. This enables extending the simulation results to much larger volumes such that observational predictions become possible. To achieve this goal, we develop an empirically calibrated star formation model based the the \emph{Renaissance} results. We include both Pop III and metal-enriched star formation, but the emphasis is on metal-enriched since the halos hosting early Pop III star formation are not well resolved. Our empirical calibration includes a number of parameters such as metal-enriched starburst mass, quiescent period following a burst, etc., that are measured directly from the simulations. This model and parameterization can easily be applied to other semi-analytic models or analytic calculations in the future.

This study is related to that of \cite{2023mccaffrey} which calibrates a semi-analytic model to \emph{Renaissance}. We note that in this case, only the largest galaxy in \emph{Renaissance} was considered and the simulation datasets used were unable to resolve the formation of Pop III. As explained in detail below, we find that while our prescription is calibrated to a region at the mean density of the Universe, it can reproduce the star formation rate density typically within a factor of $\sim$2 without any calibration on such a region. Thus, we have shown that a calibration framework such as that developed in this paper can extend the reach of numerical simulations both in volume and range of cosmic time.

This paper is structured as follows. In Section \ref{sec:renaissance}, we review the main features of the \emph{Renaissance} simulation suite. In Section \ref{sec:calibration}, we describe our parameterization of metal-enriched and Pop III star formation in \emph{Renaissance} along with our sample. In Section \ref{sec:mod_sam}, we present the calibration of a semi-analytic model to \emph{Renaissance} along with our results for metal-enriched and Pop III star formation. Finally, we summarize the results and discuss our main conclusions in Section \ref{sec:discussion}. Throughout we assume a cosmology consistent with \emph{Renaissance} using parameters from the WMAP7 $\Lambda$CDM+SZ+LENS best fit \citep{2011komatsu} with $\Omega_{\mathrm{M}} = 0.266$, $\Omega_{\mathrm{\Lambda}} = 0.734$, $\Omega_{\mathrm{b}} = 0.0449$, $h = 0.71$, $\sigma_{\mathrm{8}} = 0.81$, $n = 0.963$.

\section{The Renaissance Simulations} \label{sec:renaissance}

The \emph{Renaissance Simulations} are cosmological hydrodynamics simulations that were performed with the adaptive mesh refinement code \textsc{enzo} \citep{2014Enzo,2019Enzo} to simulate the formation of the first stars and galaxies. Radiative transfer from ionizing radiation is calculated with the \textsc{moray} package \citep{2011wise} along with chemistry and cooling for nine species of hydrogen and helium \citep{1997abel} and metallicity dependent cooling tables \citep{2009smith}.

\emph{Renaissance} consists of three different regions taken from a comoving volume of $(40 \ \mathrm{Mpc})^{3}$. The three regions named ``RarePeak", ``Normal", and ``Void" are in overdense, average, or underdense regions and evolved to redshifts of $z = 15, \ 12.5, \ \mathrm{and} \ 8$. RarePeak has a comoving volume of $(3.8 \times 5.4 \times 6.6) \ \mathrm{Mpc}^{3}$ and Normal/Void have comoving volumes of $(6.0 \times 6.0 \times 6.125) \ \mathrm{Mpc}^{3}$, respectively \citep{2015oshea}. Our semi-analytic model calibration is focused on the Normal region and then applied to RarePeak and Void for comparison.

It is important to emphasize key aspects of the metal-enriched and Population III star formation prescriptions in \emph{Renaissance} due to their importance in our star formation calibration. The metal-enriched star clusters are formed from molecular clouds with a typical radius of $\sim 6$ pc. The clusters have a total stellar mass populated with stars assuming a Salpeter IMF and emit 6000 hydrogen ionizing photons per stellar baryon over 20 Myr (the maximum lifetime of an OB star). After 4 Myr, the clusters generate $6.8 \times 10^{48} \ \mathrm{erg} \ \mathrm{M_{\mathrm{\odot}}^{-1}}$ for SNe feedback. For Pop III, individual stars are formed before the metal-enriched star clusters with a mass randomly sampled from a power-law initial mass function similar to a Salpeter IMF with an exponential decrease below a characteristic mass of $M_{\mathrm{char}} = 40 M_{\mathrm{\odot}}$. Stellar lifetimes along with Lyman-Werner and hydrogen ionizing photon luminosities are drawn from \cite{2002popIII_properties}. Stars in the mass ranges of $11 - 20 \ M_{\mathrm{\odot}}$, $20 - 40 \ M_{\mathrm{\odot}}$, and $140 - 260 \ M_{\mathrm{\odot}}$ explode as Type II supernovae and hypernovae \citep{2006nomoto}, and pair-instability supernovae \citep{2003heger}, respectively with energies exceeding $10^{51} \ \mathrm{erg}$. Stars with masses outside these ranges directly collapse into a black hole without an explosion. Further details can be found in \cite{2012wise_galaxy_birth_2}.

High-redshift galaxy observations from the CEERS \citep{2023finkelstein} and JADES surveys \citep{2023eisenstein} imply that star formation rates are likely higher than predicted by previous simulations \citep{2023arrabal,2023bunker}. \cite{2023mccaffrey} recently demonstrated that early \emph{Renaissance} galaxies typically experience rapid star formation, with a specific star formation rate of $\sim 10^{-8} \ \mathrm{yr}^{-1}$ which is consistent with the galaxies observed by CEERS and JADES. This agreement suggests that \emph{Renaissance} is a good choice for calibration to high-$z$ galaxy formation.

\section{Calibration to Renaissance} \label{sec:calibration}

\subsection{Overview} \label{subsec:cal_overview}

\begin{figure*}
\epsscale{0.65}
\plotone{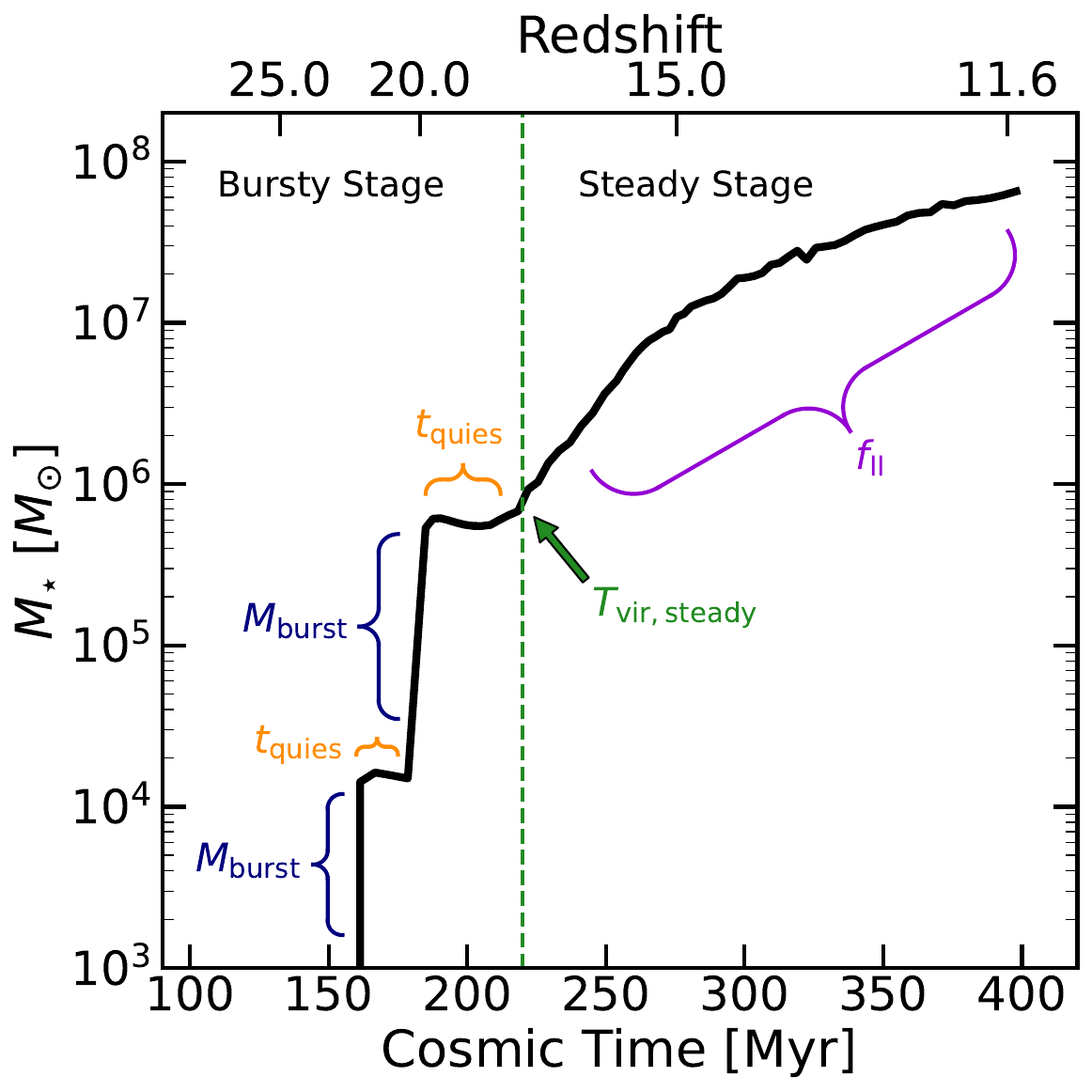}
\caption{Metal-enriched star formation history for a single \emph{Renaissance} halo. $M_{\mathrm{\star}}$ is the total mass of metal-enriched stars formed in the halo. This halo experiences two episodes of bursty star formation followed by a stage of steady star formation once the halo reaches a virial temperature threshold. The burst stellar mass, quiescent period, ionization feedback threshold temperature, and steady star formation efficiency are labeled $M_{\mathrm{burst}}$, $t_{\mathrm{quies}}$, $T_{\mathrm{vir,steady}}$, and $f_{\mathrm{II}}$ respectively. Bursty and steady stages are found throughout the sample of halos we analyze. \label{fig:calibration_overview}}
\end{figure*}

We examined star formation histories for a wide selection of \emph{Renaissance} halos, building up a representative sample detailed in Section \ref{subsec:sample}. In Figure \ref{fig:calibration_overview}, we show the metal-enriched star formation history for the most massive halo in the Normal region. Stellar evolution in this halo is representative of trends typically found in other \emph{Renaissance} halos. We find after the first generation of stars, star formation proceeds through two distinct stages, a bursty stage characterized by rapid star formation followed by a steady stage with stars forming at a more constant rate. The division of metal-enriched star formation into bursty and steady stages is the foundation for our calibration to \emph{Renaissance} and later implementation in our semi-analytic model.

The bursty stage begins with the first instance of metal-enriched star formation following the death of Pop III stars and dispersal of their metals into the interstellar medium (ISM). We do not include external enrichment in this analysis, discussed further in Section \ref{sec:mod_sam}. Forming in low-mass halos at high redshift, metal-enriched stars form in a rapid starburst at the onset of the bursty stage. Supernovae feedback from stars formed in the burst disperses any dense clumps and expels most of the gas from the shallow gravitational potential well. Following the expulsion of gas, star formation enters an extended quiescent period as gas falls back into the halo and dense clumps begin to reform. A halo can experience multiple bursts and quiescent periods dependent on its virial mass, see Section \ref{subsubsec:findings_popII} for more detail.

After the initial bursty stage, a change in star formation occurs as the steady stage begins. This transition corresponds to a fixed virial temperature which is consistent with the halo mass below which reionization has been shown to suppress star formation as a result of gas photoheating \citep[e.g.,][]{1994shapiro,1996thoul,1998gnedin,2000gnedin,2004dijkstra,2006hoeft,2008okamoto,2013sobacchi,2014noh}. Once the steady stage begins, the halo is sufficiently massive to retain enough gas to fuel continuous future star formation. Total metal-enriched star formation produced during the steady stage eventually completely dominates at lower redshifts. More detail about each component of the two stage bursty/steady model shown in Figure \ref{fig:calibration_overview} is covered in Section \ref{subsec:findings}.

\subsection{Sample} \label{subsec:sample}

We quantified trends in star formation history by focusing on the evolution of individual dark matter halos in \emph{Renaissance}. Halos grow through a combination of accretion and mergers with other halos, tracked using the halo finder \textsc{rockstar} \citep{2013behroozi_a} and merger trees generated with \textsc{consistent trees} \citep{2013behroozi_b}. The main progenitor halos in these trees represent the most massive progenitor at each step going back in time. We preferentially select halos with smoother merger histories that typically have uncomplicated star formation histories. We leave modeling of more violent mergers for future work. Halos that experience large virial mass decreases in their growth histories due to complicated mergers or other factors were excluded for our sample.

\begin{figure}
\epsscale{1.15}
\plotone{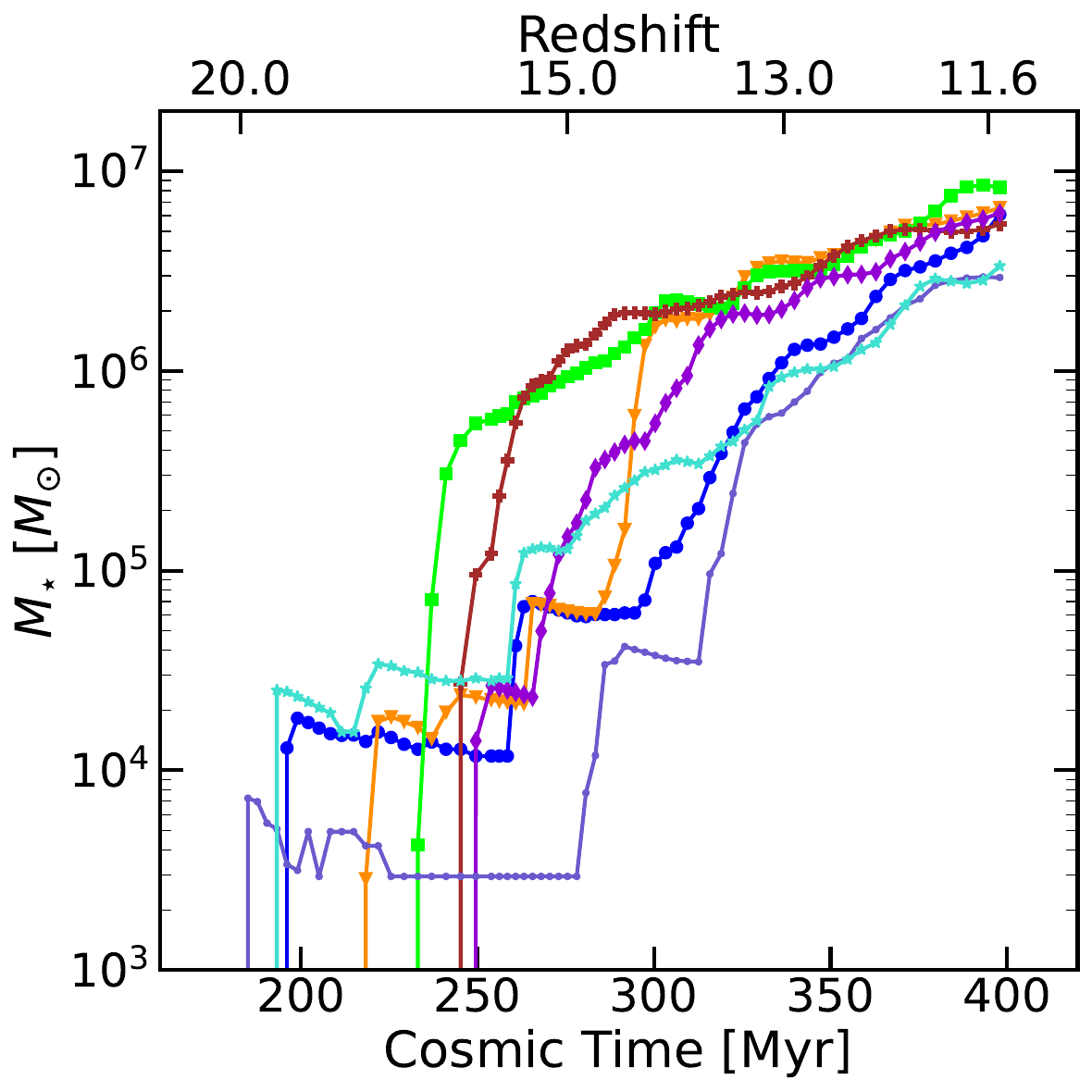}
\caption{Metal-enriched star formation histories for seven \emph{Renaissance} halos in our sample. Halos can undergo one or multiple bursts and can also skip the bursty stage, instead exhibiting star formation consistent with the steady stage (green and brown). \label{fig:sample_sf}}
\end{figure}

To calibrate the semi-analytic model, we gathered a sample of 27 dark matter halo and stellar mass histories illustrative of typical metal-enriched star formation in \emph{Renaissance} halos. For each halo in our sample, we visually identify when every starburst and quiescent period occurs along with the start of steady star formation. In Figure \ref{fig:sample_sf}, we show metal-enriched star formation histories for seven different halos. A halo typically progresses through a bursty stage with intermittent star formation followed by a steady stage, but can sometimes bypass the bursty stage entirely if star formation begins after the halo exceeds $T_{\mathrm{vir,steady}}$.

\subsection{Star Formation Parameterization} \label{subsec:findings}

Our two-stage metal-enriched star formation model has four key physical quantities. In the bursty stage this includes the mass formed in a starburst $M_{\mathrm{burst}}$, and a quiescent period following a burst $t_{\mathrm{quies}}$. The steady stage begins once the virial temperature threshold, $T_{\mathrm{vir}}$, is exceeded, signalling the start of the steady stage. Steady star formation then proceeds at a constant efficiency $f_{\mathrm{II}}$.

We focus our calibration on three different aspects of Pop III formation in \emph{Renaissance}. When Pop III stars form inside halos, the total Pop III stellar mass produced once formation conditions are met, and how much time must pass after the formation of Pop III before the formation of metal-enriched stars.

Overall, we focus on connecting star formation to halo properties such as the virial mass and do not generally consider environmental properties with the exception of feedback from cosmic reionization. However, as we discuss below in Section \ref{subsec:total_SF}, environmental properties such as the large-scale overdensity or different physics in massive halos could impact star formation. A summary of all model parameters along with best fit values discussed in the following sections can be found in Table \ref{tab:overview_table}.

\subsubsection{Metal-Enriched Star Formation} \label{subsubsec:findings_popII}

For the bursty stage, we evaluated multi-parameter log scale power law fits using least squares regression from the \textsc{scipy} curve-fit package for the burst masses and quiescent periods in our sample. We also determine standard deviations to account for the distributions of $M_{\mathrm{burst}}$ and $t_{\mathrm{quies}}$.

\begin{figure*}
\epsscale{0.85}
\plotone{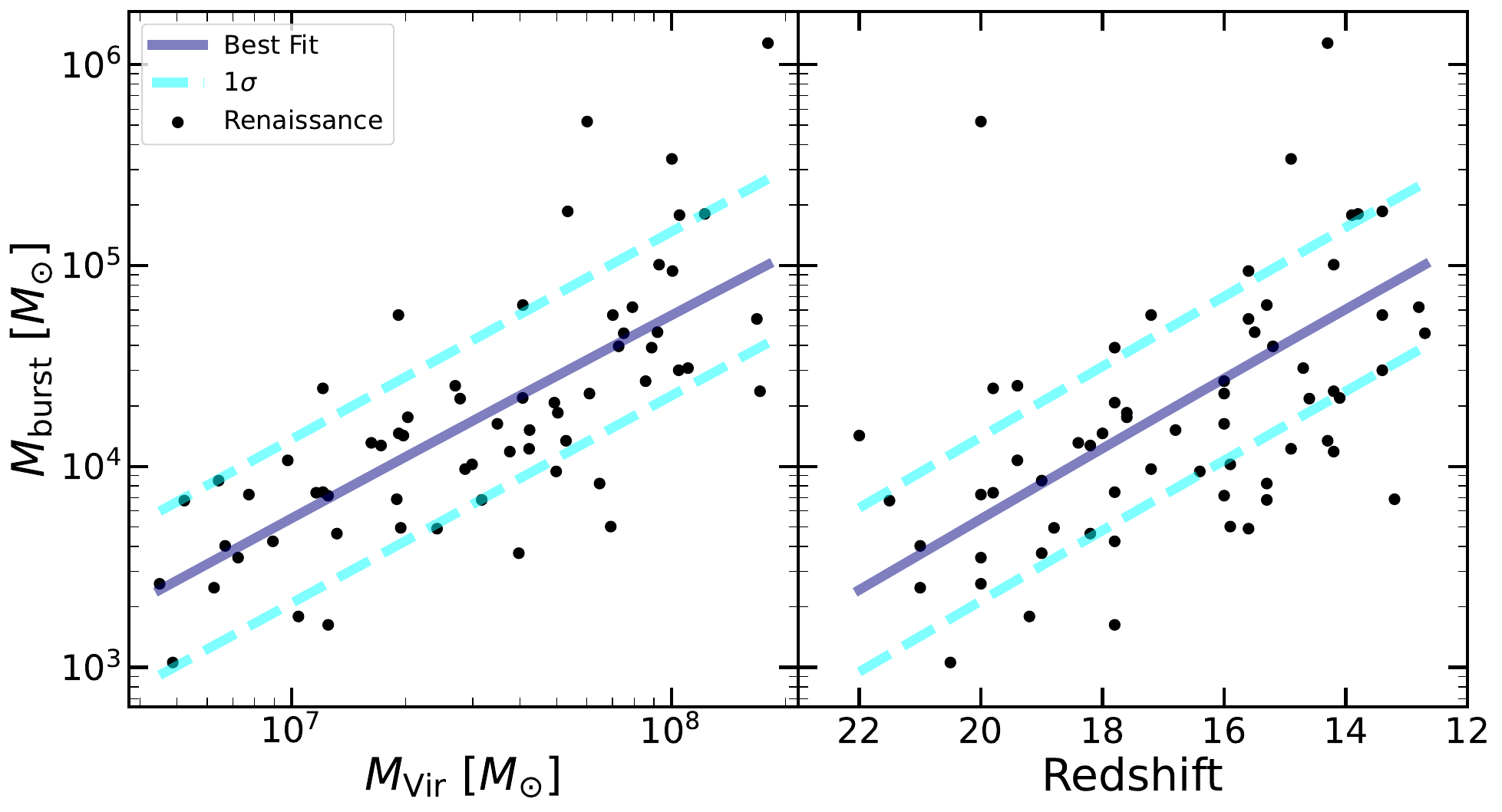}
\caption{Metal-enriched starburst masses from the sample of \emph{Renaissance} halos during the bursty stage. Best fit two parameter power law (solid-blue) along with 1$\sigma$ lines (dashed-cyan). Burst mass increases for larger mass halos and at lower redshifts. \label{fig:burst_mass}}
\end{figure*}

In Figure \ref{fig:burst_mass}, we find that the metal-enriched burst masses in our sample are closely correlated with both the halo virial mass and the redshift of the burst. We parameterize the stellar mass formed in a burst as

\begin{equation}
M_{\mathrm{burst}} = \alpha \left( \frac{M_{\mathrm{vir}}}{10^{7} M_{\mathrm{\odot}}} \right)^{\gamma} \left( \frac{1 + z}{20} \right)^{\delta} M_{\mathrm{\odot}}
\label{eqn:burst_mass_eqn}
\end{equation}
where $M_{\mathrm{vir}}$ and $z$ are respectively the halo virial mass and redshift before a starburst. The best fit parameters are $\alpha = 5.364 \times 10^{3}$, $\gamma = 1.092$, and $\delta=0.578$. The distribution of the sample around $M_{\mathrm{burst}}$ is approximately log-normal with a standard deviation of log$_{\mathrm{10}}\left( M_{\mathrm{burst}} / M_{\mathrm{\odot}} \right) = 0.408$. As halos reach higher masses at lower redshift, the stellar mass of metal-enriched stars formed in a burst increases. 

\begin{figure}
\epsscale{1.15}
\plotone{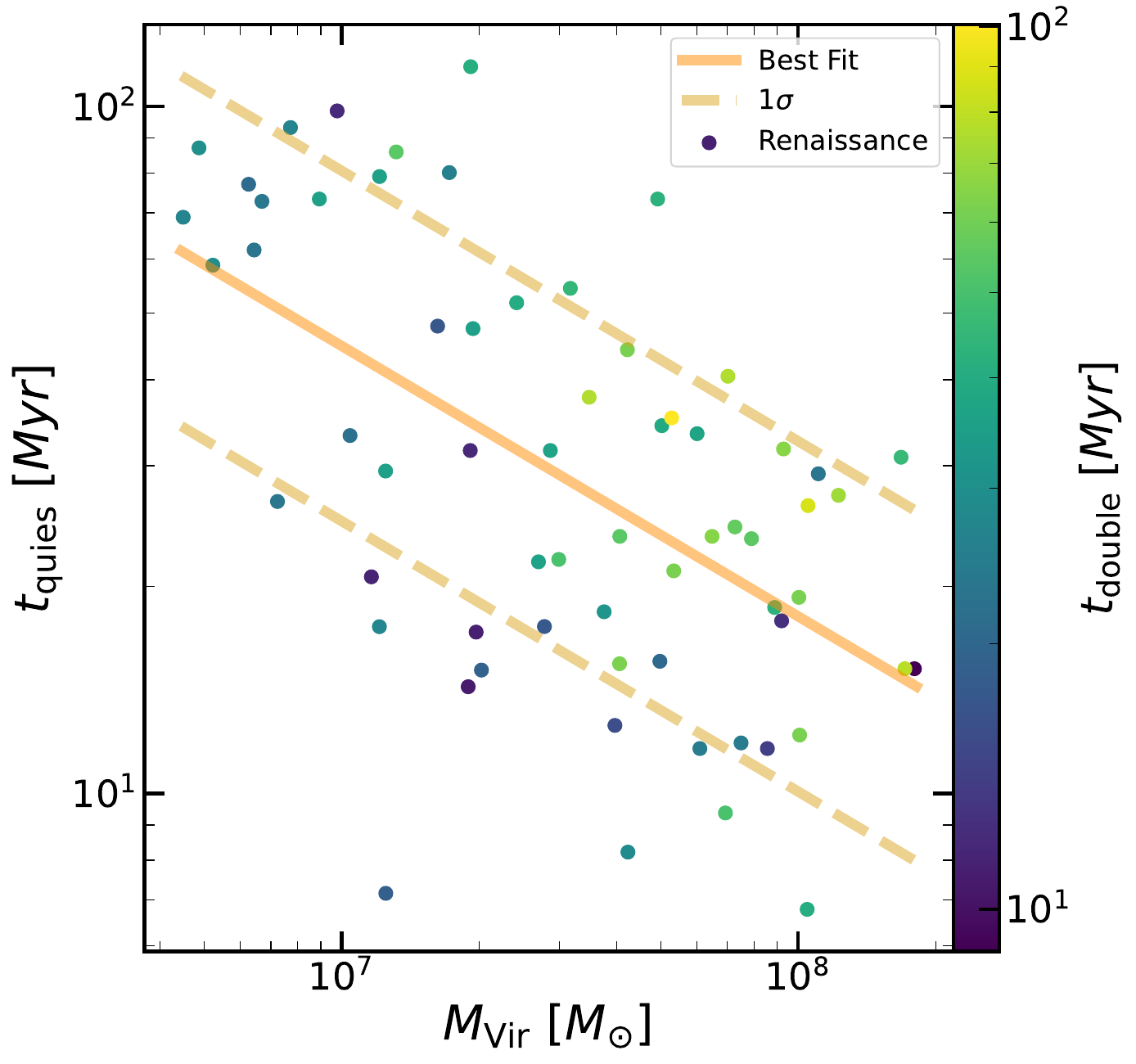}
\caption{Quiescent times from the sample of \emph{Renaissance} halos during the bursty stage. Power law best fit (solid-orange) and 1$\sigma$ lines (dashed-gold). The color bar values correspond to the time in millions of years it takes the halo to double in virial mass. Overall the quiescent time between star formation events decreases for higher mass halos. \label{fig:delay_time}}
\end{figure}

For the quiescent periods following a starburst, we find a strong negative correlation with the halo virial mass, as shown in Figure \ref{fig:delay_time}. The best fit is

\begin{equation}
t_{\mathrm{quies}} = \epsilon \left( \frac{M_{\mathrm{vir}}}{10^{7} M_{\mathrm{\odot}}} \right)^{\kappa} \text{Myr}
\label{eqn:quiescent_period}
\end{equation}
where $M_{\mathrm{vir}}$ is the virial mass following a starburst and the best fit parameters are $\epsilon = 4.481$ and $\kappa = -0.394$. The distribution of the sample around $t_{\mathrm{quies}}$ is approximately log-normal with a standard deviation of log$_{\mathrm{10}}\left( t_{\mathrm{quies}} / \mathrm{Myr} \right) = 0.255$. As halos increase in mass, building up deeper gravitational potential wells better able to hold onto gas, a typical quiescent period declines from $\sim$100 Myr to $\sim$10 Myr for halo masses between $\sim10^{7}$--$10^{8} \ M_{\mathrm{\odot}}$.

If shorter quiescent periods are related to the retention of gas reservoirs that fuels subsequent star formation, then the rate the halo is accreting mass should impact the length of the quiescent period. A fast growing halo could potentially recover gas ejected from a starburst and reduce the quiescent period before the next instance of star formation accordingly. We parameterized this effect by considering the time for a halo to double in virial mass following a starburst, $t_{\mathrm{double}}$ (see Figure \ref{fig:delay_time}). Incorporating $t_{\mathrm{double}}$ in our power law did not significantly change the overall fit, with the variance of the sample data from the fit being reduced by less than 20 percent. Therefore, we do not include $t_{\mathrm{double}}$ in the final fits.

With the start of the steady stage, we found that the transition occurs at virial temperatures within a factor of 2 of $T_{\mathrm{vir,steady}} = 1.76 \times 10^{4}$ K corresponding to an ionization feedback mass for almost all star formation histories in our sample. We assume a mean molecular weight of $\mu = 0.6$ consistent with fully ionized primordial gas for the temperature \citep{2001barkana}. Due to the small range in temperature, we define the onset of the steady stage to occur once the virial temperature exceeds $T_{\mathrm{vir,steady}}$. The virial temperature at the beginning of the steady stage showed weak or no correlation with other halo properties.

Once the steady stage is underway, we found metal-enriched star formation to be well described using a system of differential equations presented in \cite{2022furlanetto}. These coupled equations derive the evolution of both the gas and stellar mass of a halo using a simplified implementation of the `bathtub' model \citep[e.g.,][]{2010bouche,2012dave,2014dekel}:

\begin{equation}
\dot{M}_{\mathrm{*}} = \frac{f_{\mathrm{II}}}{t_{\mathrm{ff}}} M_{\mathrm{gas}},
\label{eqn:steady_star_mass_eqn}
\end{equation}

\begin{equation}
\dot{M}_{\mathrm{gas}} = \dot{M}_{\mathrm{acc}} - \dot{M}_{\mathrm{*}} - \eta_{\mathrm{SN}} \dot{M}_{\mathrm{*}},
\label{eqn:steady_gas_mass_eqn}
\end{equation}
where $\dot{M}_{\mathrm{*}}$ is the stellar mass growth rate, $\dot{M}_{\mathrm{gas}}$ is the rate of change in halo gas mass, $\dot{M}_{\mathrm{acc}}$ is the halo gas accretion rate, $f_{\mathrm{II}}$ is star formation efficiency, $t_{\mathrm{ff}}$ is the free fall time assumed to be a tenth of the Hubble time (which corresponds to the dynamical time at the viral radius of a dark matter halo described in \cite{2013loeb}), and $\eta_{\mathrm{SN}}$ is the gas ejection efficiency for supernovae feedback.

Two parameters that warrant further discussion are the SNe ejection efficiency, $\eta_{\mathrm{SN}}$, and steady star formation efficiency, $f_{\mathrm{II}}$. We determine $\eta_{\mathrm{SN}}$ using a similar prescription to \cite{2021sassano}, with a SNe ejection efficiency of

\begin{equation}
\eta_{\mathrm{SN}} = \beta \times \frac{2 E_{\mathrm{SN}} \epsilon_{\mathrm{SN}} R_{\mathrm{SN}}}{v_{\mathrm{esc}}^{2}}
\label{eqn:sn_eject_eqn},
\end{equation}
where $E_{\mathrm{SN}}$ is the average SNe explosion energy, $\epsilon_{\mathrm{SN}}$ is the SNe kinetic energy conversion fraction, $R_{\mathrm{SN}}$ is the SNe rate per mass, $v_{\mathrm{esc}}$ is the escape velocity from the halo, and $\beta$ is the SNe ejection scale factor.

We assume the average energy for $E_{\mathrm{SN}}$ is $10^{51}$ erg \citep{2006nomoto}. With most of the energy from a supernovae lost as thermal energy, only a small fraction is converted into kinetic energy coupled to the gas \citep{2015walch}. We adopt a value of $\epsilon_{\mathrm{SN}} = 1.6 \times 10^{-3}$ for this efficiency that is calibrated to observations \citep{2021sassano}. For $R_{\mathrm{SN}}$, we use metal-enriched SNe feedback energy per mass in agreement with \emph{Renaissance} of $6.8 \times 10^{48}$ erg $\mathrm{M}_{\mathrm{\odot}}^{-1}$ \citep{2012wise_galaxy_birth_1}. Dividing this quantity by $E_{\mathrm{SN}}$ yields a SNe rate of $R_{\mathrm{SN}} = 6.8 \times 10^{-3} \ \mathrm{M}_{\mathrm{\odot}}^{-1}$. Finally, $v_{\mathrm{esc}}$, is set to be the circular velocity of a halo excluding a factor of $\sqrt{2}$ in the numerator of the escape velocity using the relation $\sqrt{GM / r_{\mathrm{vir}}}$ from \cite{2001barkana}.

We lower the SNe efficiency using a SNe ejection scale factor $\beta$ calibrated to high-mass halos. Values between $\beta = 0$--$0.1$ generate star formation consistent with our sample and we find $\beta = 0.01$ produces total metal-enriched star formation that agrees well with \emph{Renaissance}. For $\beta = 1.0$, $\eta_{\mathrm{SN}}$ significantly suppresses star formation in the most massive halos, producing stellar masses around an order of magnitude lower than \emph{Renaissance} due to the efficient ejection of a large fraction of gas early in the steady stage.

To calibrate the steady star formation efficiency in Equation \ref{eqn:steady_star_mass_eqn} to \emph{Renaissance}, we determine values for $f_{\mathrm{II}}$ in agreement with our sample of metal-enriched star formation histories using the following method. First, we separate the steady stage from the bursty stage for each star formation history in our sample. We set the initial stellar mass equal to \emph{Renaissance} and assume the gas mass is consistent with the cosmic baryon fraction of the halo virial mass. Next, we evolve Equations \ref{eqn:steady_star_mass_eqn} and \ref{eqn:steady_gas_mass_eqn} over the entire steady stage and perform a regression using the \textsc{scipy} curve-fit package to determine the best fit for $f_{\mathrm{II}}$.

\begin{figure}
\epsscale{1.15}
\plotone{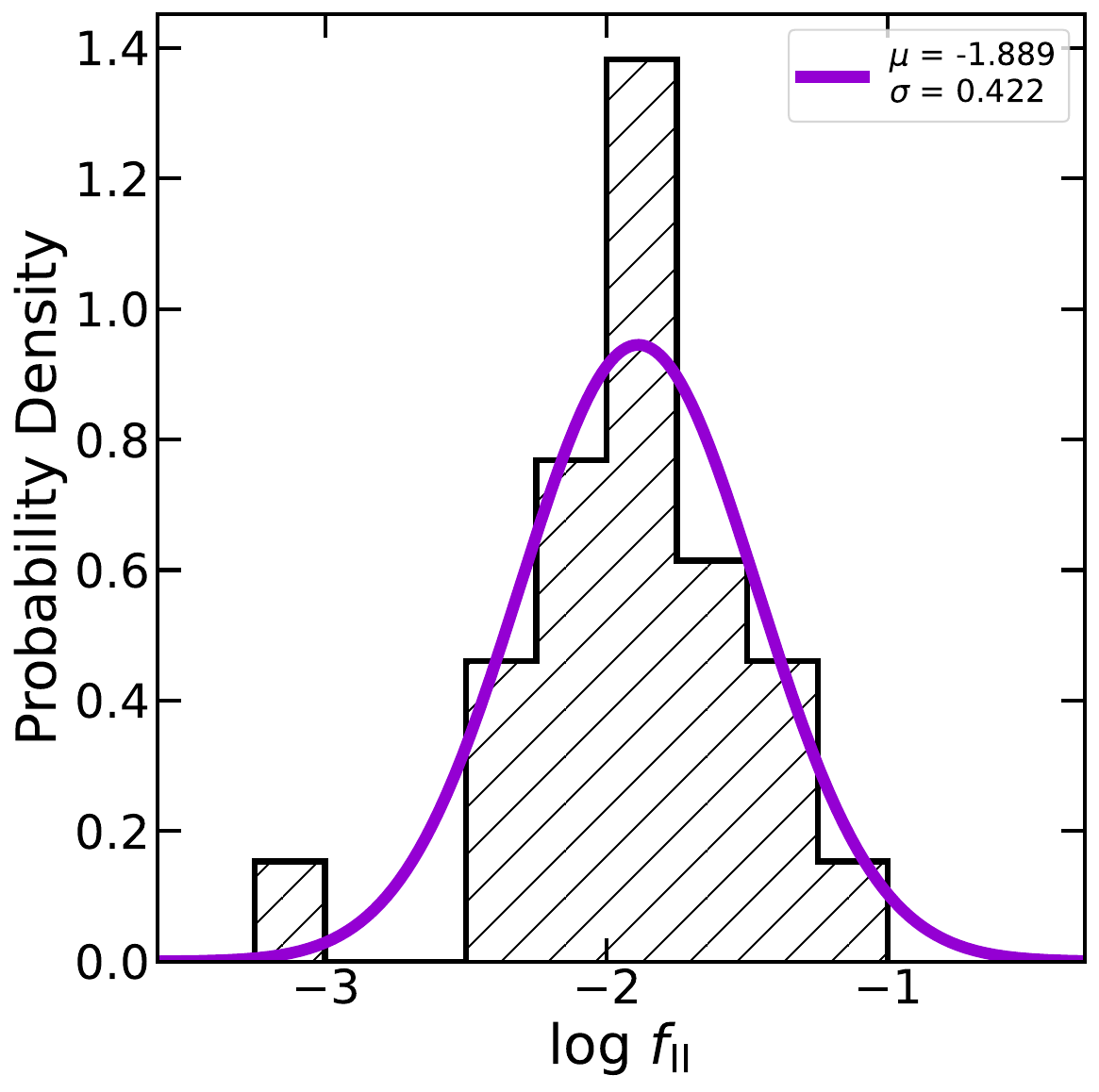}
\caption{Metal-enriched star formation efficiencies for the steady stage in \emph{Renaissance} halos. Each $f_{\mathrm{II}}$ were determined by finding the best fit for Equations \ref{eqn:steady_star_mass_eqn} and \ref{eqn:steady_gas_mass_eqn} during the steady stage of a halo in the sample. \label{fig:fII_dist}}
\end{figure}

Figure \ref{fig:fII_dist} shows the distribution of best fit steady stage star formation efficiencies. We find the distribution is roughly log-normal and centered around a mean efficiency of $f_{\mathrm{II}} = 0.013$ with a standard deviation of log $\left( f_{\mathrm{II}} \right) = 0.422$. The predicted metal-enriched stellar mass using the best fit efficiency with Equations \ref{eqn:steady_star_mass_eqn} and \ref{eqn:steady_gas_mass_eqn} at most deviates from \emph{Renaissance} by a factor of $\sim$1.5. We note that outliers such as the low efficiency of $f_{\mathrm{II}} = 0.0006$ have a large impact on the variance of the distribution due to the small sample size.

We found that a halo mass dependent $f_{\mathrm{II}}$ of the functional form $f_{\mathrm{II}} M_{\mathrm{vir}}^{\tau}$ also produced good fits for the \emph{Renaissance} star formation histories of most halos during the steady stage. However for the largest halos at low redshift, this produces unphysically high star formation efficiencies and therefore we do not include it in our model.

\subsubsection{Population III Star Formation} \label{subsubsec:findings_popIII}

Population III stars form in pristine metal-free halos that exceed the minimum mass where dense gas clouds can condense due to cooling from roto-vibrational molecular hydrogen transitions \citep{1996haiman_1,1996haiman_2,1997tegmark,2001machacek}. The $\mathrm{H_{2}}$ fraction is fundamental to determine if a halo forms Pop III stars and is included as a formation condition in \emph{Renaissance} accordingly (see Section \ref{sec:renaissance}). Therefore, physical processes that change the $\mathrm{H_{2}}$ fraction are important to consider.

Lyman-Werner photons with energies between 11.2-13.6 eV emitted by the first stars can efficiently photodissociate $\mathrm{H_{2}}$ through the two-step Solomon process \citep{1967stecher}. The LW photons travel over cosmological scales to neighboring metal-free halos, suppressing $\mathrm{H_{2}}$ cooling and the formation of Pop III stars \citep{1987dekel,1997haiman,2000haiman,2001machacek,2007wise,2008oshea,2009ahn,2014visbal}. We follow \cite{2013fialkov} to compute the minimum mass for molecular cooling that can form stars in minihalos using the relation, 

\begin{equation}
M_{\mathrm{H_{2}}} = 2.5 \times 10^{5} \left( \frac{1+z}{26} \right)^{-3/2} \times \left[1 + 6.96(4 \pi J_{\mathrm{LW,21}})^{0.47} \right] M_{\mathrm{\odot}}
\label{eqn:H2_min_mass}
\end{equation}
where the first term is the minimum halo mass in the absence of a LW background to suppress $\mathrm{H_{2}}$ determined using the same assumptions as \cite{2020visbal} and $J_{\mathrm{LW,21}}$ is the local LW background in units of $10^{-21} \ \mathrm{erg} \ \mathrm{s}^{-1} \ \mathrm{cm}^{-2} \ \mathrm{Hz}^{-1} \ \mathrm{sr}^{-1}$. We use the background presented in \cite{2016xu} that self-consistently calculates a LW background from star formation in the Normal region.

Once a halo reaches virial temperatures of $T_{\mathrm{vir,steady}} \gtrsim 10^{4} \ \mathrm{K}$, collisionally excited line emission from atomic hydrogen becomes the dominant cooling process. For the atomic cooling mass threshold we adopt the relation,

\begin{equation}
M_{\mathrm{a}} = 5.4 \times 10^{7} \left( \frac{1+z}{11} \right)^{-3/2} M_{\mathrm{\odot}}
\label{eqn:atomic_cooling_mass}
\end{equation}
which is determined using the cosmological hydrodynamical simulations of \cite{2014fernandez}.

The \emph{Renaissance} simulations have a dark matter mass resolution of $2.9 \times 10^{4} M_{\mathrm{\odot}}$ and the halo mass function is well-resolved for halos with masses $\gtrsim 2 \times 10^{6} M_{\mathrm{\odot}}$ \citep{2015oshea}. This artificially sets a lower-limit on the minimum halo mass scale for Pop III star formation. Simulations by \cite{2021kulkarni} and \cite{2021schauer} find minimum masses of order $10^{5}$-$10^{6} M_{\mathrm{\odot}}$ at high redshift for similar LW backgrounds which are unable to be resolved by \emph{Renaissance}. Therefore, we adopt a mass resolution limit of $7 \times 10^{6} M_{\mathrm{\odot}}$ below which Pop III is assumed to be unable to form. This allows our model to closely reproduce Pop III star formation by delaying star formation in unresolved low mass halos similarly to \emph{Renaissance}.

\begin{figure*}
\epsscale{0.98}
\plotone{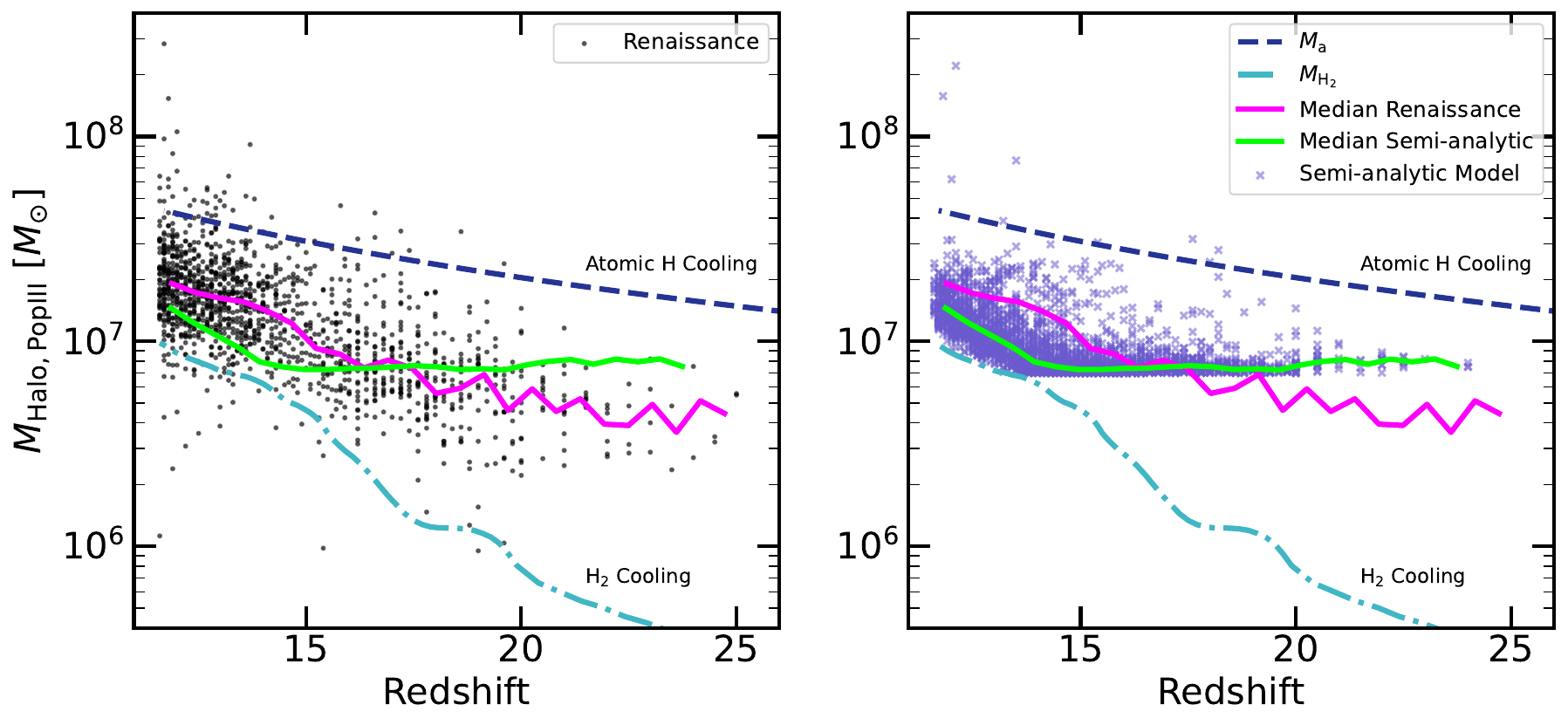}
\caption{Dark matter masses of halos that host the first Population III star formation for Renaissance (left panel) and one realization of the semi-analytic model (right panel). We show the median halo mass hosting Pop III for Renaissance (magenta) and the semi-analytic model (green). The dashed and dot-dashed curves show $M_{\mathrm{a}}$ and $M_{\mathrm{H_{2}}}$, respectively. The sharp cut-off in halo mass for the semi-analytic model in the right panel is due to the mass resolution limit of $7 \times 10^{6} M_{\mathrm{\odot}}$. \label{fig:first_popIII}}
\end{figure*}

The left panel of Figure \ref{fig:first_popIII} shows the dark matter halo masses of \emph{Renaissance} halos that host the first Population III star formation, $M_{\mathrm{Halo,PopIII}}$, as a function of redshift. We include $M_{\mathrm{a}}$ and $M_{\mathrm{H_{2}}}$ using a $J_{\mathrm{LW,21}}$ equivalent to the global LW background (a larger local value for $J_{\mathrm{LW,21}}$ than the background would result in a higher $M_{\mathrm{H_{2}}}$).

To produce Pop III stellar masses consistent with \emph{Renaissance}, we determined the total mass of Pop III stars formed during the initial instance of star formation in \emph{Renaissance} halos. We found that the total Pop III mass, $M_{\mathrm{Total,PopIII}}$, closely follows a log-normal distribution centered on a mass of $\approx 150 M_{\mathrm{\odot}}$ with a standard deviation of log$_{\mathrm{10}}\left( M_{\mathrm{Total,PopIII}} / M_{\mathrm{\odot}} \right) = 0.371$. Both the virial mass of the halo and the redshift were found to have minimal correlation with total Pop III mass formed.

\begin{figure}
\epsscale{1.15}
\plotone{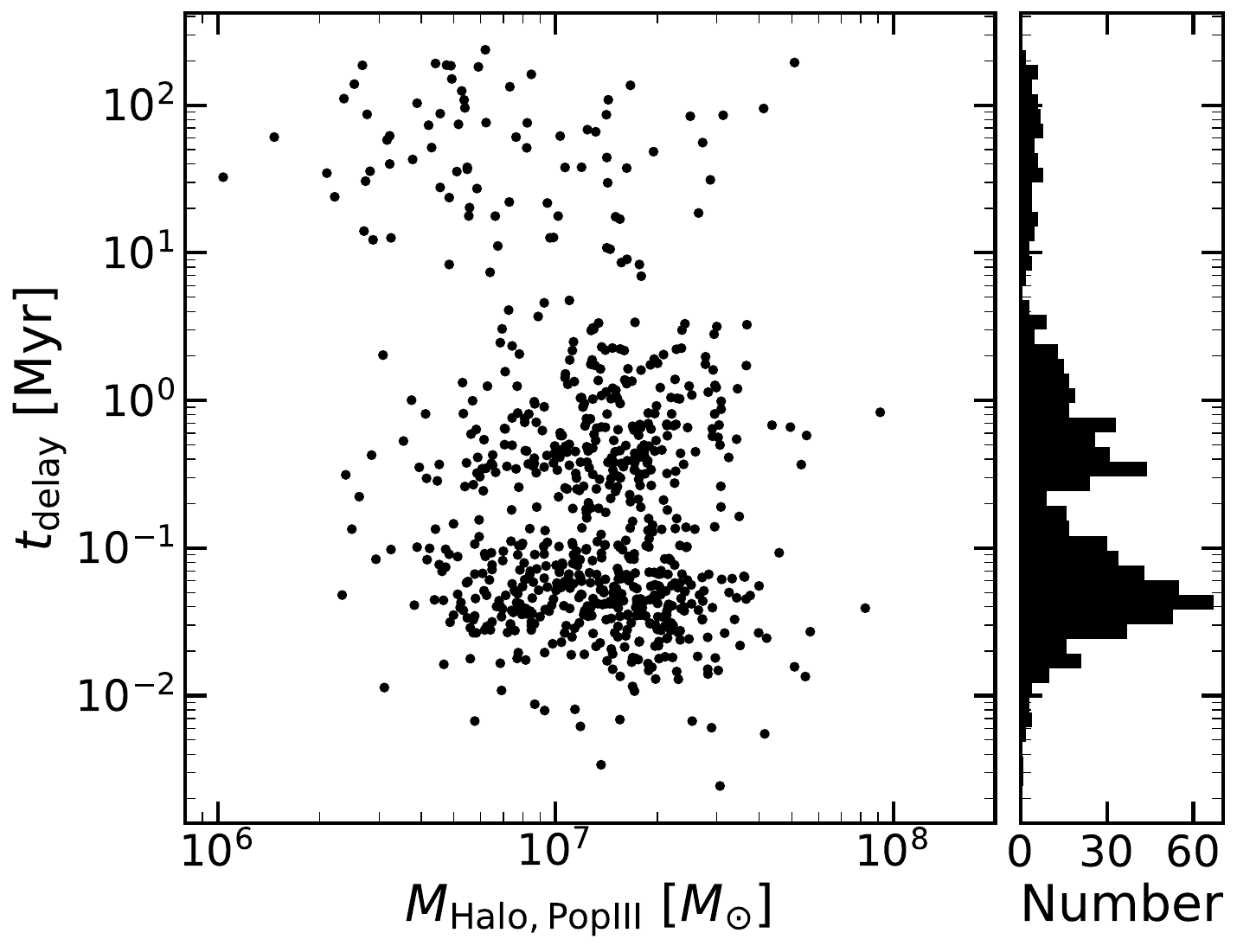}
\caption{The delay period before the formation of metal-enriched stars following the death of the first Pop III star compared with the dark mass halo mass at formation of Pop III. \label{fig:popIII_to_popII_delay}}
\end{figure}

Finally, we calculated the typical delay between the first Pop III supernovae in a halo, occurring after a mass-dependent stellar lifetime from \cite{2002popIII_properties}, with the beginning of metal-enriched star formation. Figure \ref{fig:popIII_to_popII_delay} shows the distribution of the delay times as a function of halo mass. We find three distinct populations of delay times. One population has metal-enriched star formation starting rapidly between 0.01--0.1 Myr after the first Pop III supernova. Another population has delays between 0.1--3 Myr and a final smaller group has longer delay times of 10--100 Myr. We investigated deeper potential wells in high mass halos retaining metal-enriched gas, ionizing feedback suppressing star formation, and powerful Pop III supernovae as physical origins for these three populations. However, the delay times show little to no correlation with halo mass, virial temperature, and total supernovae energy emitted by Pop III stars before the onset of metal-enriched formation.

The halos that begin to form metal-enriched stars after less than a million years are of particular interest due to the short timescales for metals to be dispersed into the nearby ISM. In this case, star formation is triggered in nearby clumps that have survived the radiative phase of the Pop III stars. A SN blastwave will compress the clumps above the density threshold for star formation, however, some of these cases could be artificial. It is possible a number of these clumps meet the criteria for star formation in the simulation, but are not gravitationally bound and should instead dissipate sometime later. The typical \emph{Renaissance} output cadence of $\sim 4$ Myr hinders further investigation into these clumps and the underlying cause of the quick transition to metal-enriched star formation. To be as consistent with \emph{Renaissance} as possible, we consider a $t_{\mathrm{delay}}$ ranging between 0.01--100 Myr with the understanding that numerical issues could be responsible for the shortest delays.

\begin{deluxetable*}{lccc}
\tabletypesize{\scriptsize}
\tablewidth{0pt} 
\tablecaption{Parameters for Star Formation in the Semi-analytic Model \label{tab:overview_table}}
\tablehead{
\colhead{Parameter} & \colhead{Description} & \colhead{Fit Values} & \colhead{Standard Deviation}
} 
\startdata 
$M_{\mathrm{burst}}$ & Metal-enriched starburst mass & $\alpha = 5.364 \times 10^{3}$ & $0.408$\\
{       } & {       } & $\gamma = 1.092$ & {       }\\
{       } & {       } & $\delta = 0.578$ & {       }\\
\hline
$t_{\mathrm{quies}}$ & Quiescent period following metal-enriched starburst & $\epsilon = 4.481$ & $0.255$\\
{       } & {       } & $\kappa = -0.394$ & {       }\\
\hline
$f_{\mathrm{II}}$ & Star formation efficiency during steady stage & $0.013$ & $0.422$\\
\hline
$M_{\mathrm{Total,PopIII}}$ & Total Pop III Stellar Mass formed after halo exceeds $M_{\mathrm{min}}$ & $150$ $M_{\mathrm{\odot}}$& $0.371$\\
\hline
$t_{\mathrm{delay}}$ & Delay before metal-enriched star formation & $0.01-100$ Myr & ...\\
\enddata 
\tablecomments{$M_{\mathrm{burst}}$ is given by Equation \ref{eqn:burst_mass_eqn}. 
$t_{\mathrm{quies}}$ is given by Equation \ref{eqn:quiescent_period}.
$f_{\mathrm{II}}$ is given by the best fit metal-enriched star formation efficiencies in Figure \ref{fig:fII_dist}.
$M_{\mathrm{Total,PopIII}}$ is described in Section \ref{subsubsec:findings_popIII}. $M_{\mathrm{burst}}$, $t_{\mathrm{quies}}$, $f_{\mathrm{II}}$, and $M_{\mathrm{Total,PopIII}}$ each follow a log-normal distribution. The last column shows the standard deviation of the log$_{\mathrm{10}}$ of the corresponding quantity as mentioned in the text. $t_{\mathrm{delay}}$ is described in Section \ref{subsubsec:findings_popIII} and follows a log-uniform distribution.}
\end{deluxetable*}

\section{Model Modifications and Results} \label{sec:mod_sam}

We modify the semi-analytic model introduced in \cite{2020visbal} which is based on dark matter halo merger trees from cosmological N-body simulations, and incorporates star formation at a constant efficiency, feedback from Lyman-Werner radiation and growth of H \textsc{ii} regions in the intergalactic medium. These processes are calculated on a grid using fast-fourier transforms (FFTs) using 3D spatial positions and clustering of halos.

Star formation in the updated semi-analytic model consists of Pop III formation followed by two stages of metal-enriched star formation with a bursty and steady stage. The calibration to \emph{Renaissance} is implemented by including a minimum mass for Pop III star formation, $M_{\mathrm{min}}$, along with total Pop III stellar masses and delay periods before the beginning of metal-enriched star formation that are consistent with \emph{Renaissance}. Additionally, for metal-enriched star formation we introduce our findings for $M_{\mathrm{burst}}$, $t_{\mathrm{quies}}$, $T_{\mathrm{vir,steady}}$, and $f_{\mathrm{II}}$ from the bursty and steady stages previously presented in Section \ref{subsubsec:findings_popII}. See Table \ref{tab:overview_table} for a concise overview of the implementation into the model. Finally, we do not include external enrichment in our calibrations. External enrichment of neighboring star-forming halos was found to marginally increase the metal-enriched SFRD and slightly reduce the Population III SFRD in \cite{2020visbal}. While this does not preclude external enrichment being more important in dense, highly-clustered environments, for now we leave calibration to external enrichment in \emph{Renaissance} for future work.

Now we will cover the modifications to the \cite{2020visbal} model in more detail. We assume the minimum mass for Pop III star formation to be the minimum of $M_{\mathrm{H_{2}}}$ and $M_{\mathrm{a}}$ with a mass resolution lower-limit of $\gtrsim 7 \times 10^{6} M_{\mathrm{\odot}}$ from Section \ref{subsubsec:findings_popIII}. For the local LW background $J_{\mathrm{LW,21}}$ term in $M_{\mathrm{H_{2}}}$, an external background is included in addition to the local component from Pop III and metal-enriched stellar mass. We implement a background that self-consistently calculates a LW background from star formation in the Normal region discussed in Section \ref{subsubsec:findings_popIII}. If a halo with no prior star formation is located in an ionized region, we increase the minimum mass to $M_{\mathrm{min}} = 1.5 \times 10^{8} \left( \frac{1+z}{11} \right)^{-3/2} M_{\mathrm{\odot}}$ (in agreement with \cite{2004dijkstra}).

The right panel of Figure \ref{fig:first_popIII} shows $M_{\mathrm{Halo,PopIII}}$ predicted by our model. There are two differences between our model and \emph{Renaissance} that are important to highlight. First, values of $M_{\mathrm{Halo,PopIII}}$ in our model are much more tightly clustered at lower halo masses than \emph{Renaissance}. A potential consequence of this behavior is the same halo in our model sometimes can form Pop III at a lower halo mass corresponding to an earlier time. Second, $M_{\mathrm{Halo,PopIII}}$ in \emph{Renaissance} is typically smaller for $z>17$ and higher for $z<17$ compared to the semi-analytic model. Additionally, $M_{\mathrm{Halo,PopIII}}$ displays greater variability for \emph{Renaissance}. The consequences of these differences will be further discussed in Section \ref{subsec:total_SF}.

Once the mass of a halo exceeds $M_{\mathrm{min}}$, we use the findings for Pop III formation discussed in Section \ref{subsubsec:findings_popIII}. First, we randomly sample a total Pop III stellar mass from the log-normal \emph{Renaissance} distribution centered on $\approx 150 \ M_{\mathrm{\odot}}$. Then, we assign a delay period before metal-enriched star formation can begin, $t_{\mathrm{delay}}$, that ranges between 0.01--100 Myr. 

After $t_{\mathrm{delay}}$ following Pop III star formation has elapsed, we determine if a halo is the bursty or steady stage by comparing its virial temperature to the $T_{\mathrm{vir,steady}} = 1.76 \times 10^{4}$ K threshold. If in the bursty stage, we determine the total stellar mass formed in a starburst by randomly sampling from a log-normal distribution centered on $M_{\mathrm{burst}}$ from Equation \ref{eqn:burst_mass_eqn}. A quiescent period sampled from another log-normal distribution centered on $t_{\mathrm{quies}}$ from Equation \ref{eqn:quiescent_period} is assigned to the halo and its descendants. The halo will then either experience a subsequent starburst after the quiescent period has passed or it will exceed the $T_{\mathrm{vir,steady}}$ threshold and enter the steady star formation stage.

At the beginning of the steady stage, we initially assume the halo contains gas consistent with the cosmic baryon fraction of its virial mass. A star formation efficiency, $f_{\mathrm{II}}$, is then randomly sampled from the roughly log-normal distribution in Figure \ref{fig:fII_dist}. This efficiency will be passed to all descendants of this halo unless a merger with other halos in the steady stage occurs. Then, the average of progenitor halo star formation efficiencies is assigned to the descendant instead. We calculate metal-enriched star formation and follow the subsequent evolution of gas mass using Equations \ref{eqn:steady_star_mass_eqn} and \ref{eqn:steady_gas_mass_eqn} respectively.

\subsection{Individual Star Formation Histories} \label{subsec:individual_SFHs}

\begin{figure*}
\epsscale{0.85}
\plotone{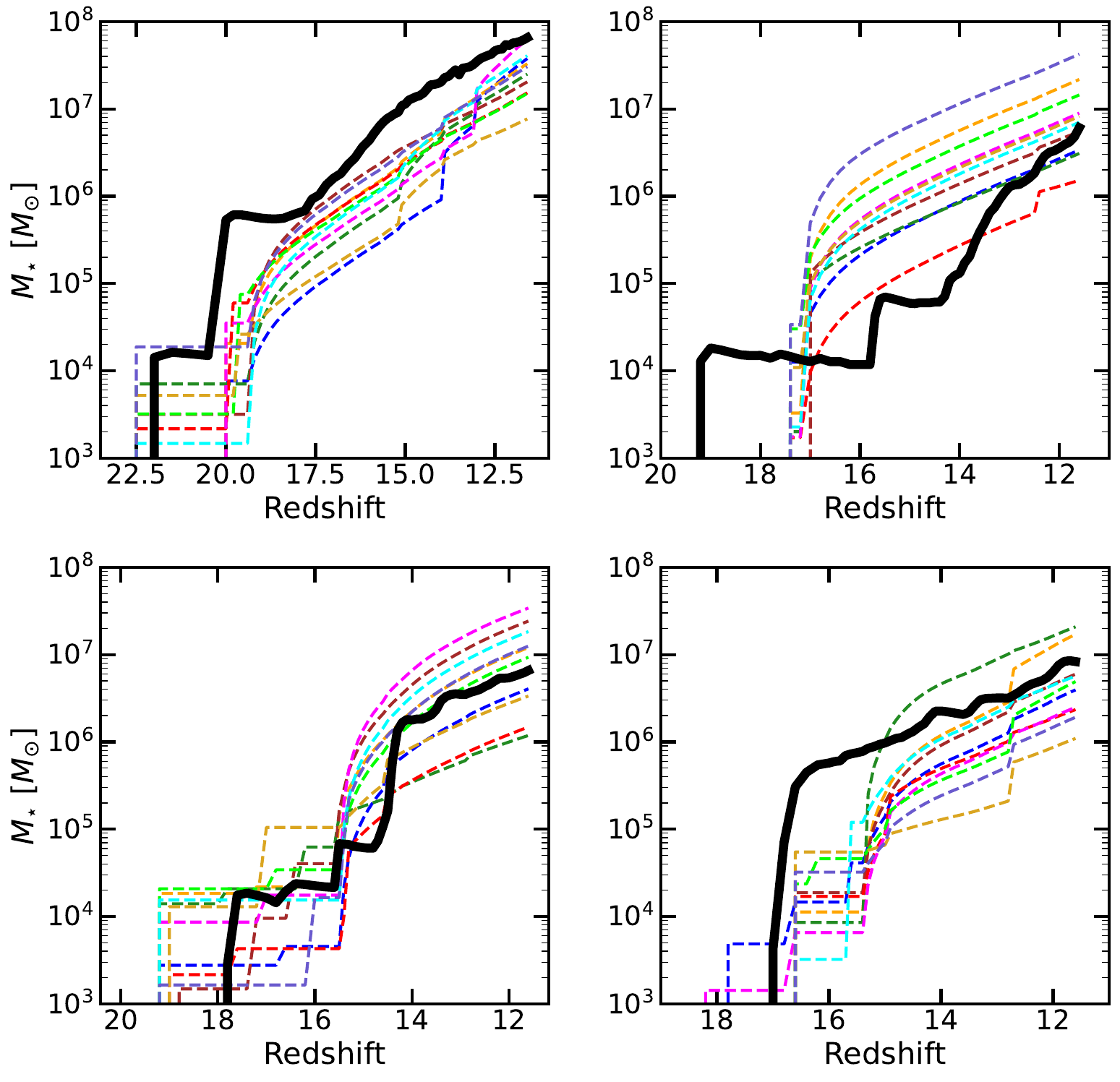}
\caption{Total metal-enriched stellar masses and redshifts of four different main progenitor halos. The \emph{Renaissance} metal-enriched star formation history (black) is compared with ten different realizations of the semi-analytic model (dashed). \label{fig:SAM_single_SFH}}
\end{figure*}

To determine the effectiveness of our calibration, we compare different realizations of our model with \emph{Renaissance} for metal-enriched halo star formation histories. We begin by running 10 different random seeds of our semi-analytic model on the \emph{Renaissance} dark matter halo merger trees. Figure \ref{fig:SAM_single_SFH} shows a selection of metal-enriched SFHs from our simulations. The top-left panel shows the metal-enriched SFH for the most massive halo in the \emph{Renaissance} Normal region. The large second starburst and high $f_{\mathrm{II}} \approx 0.1$ are extreme values in our distributions for $M_{\mathrm{burst}}$ and $f_{\mathrm{II}}$, making them correspondingly difficult to reproduce in our model. The SFHs in the top-right and bottom-right panels transition from the bursty to the steady stage differently from \emph{Renaissance}. Our fixed threshold for the transition at $T_{\mathrm{vir,steady}} = 1.76 \times 10^{4}$ K can lead to more starbursts or an earlier transition to the steady stage, by the final simulation redshift, the total star formation produced during the steady stage lies within the scatter produced by the model. The bottom-left panel shows a SFH with bursty and steady stages that fall well within the scatter of our model.

Generally, there is good agreement between star formation histories produced by the semi-analytic model and \emph{Renaissance}. By the final simulation redshift when metal-enriched stars dominate, star formation for \emph{Renaissance} halos typically lie within the range predicted by our model and agreement for the best model is within a factor of 2. Additionally, the chaotic bursty stage is also well reproduced if metal-enriched star formation begins at a similar time.

While the overall agreement with \emph{Renaissance} is considerable, the model does not perfectly capture the variance in individual star formation histories. Our choice to calibrate to a sample of main progenitor halos without violent major mergers in \emph{Renaissance} could potentially not be fully representative of all halos. Additionally, the differential equations used during the steady stage produce star formation that is smoother than \emph{Renaissance}, lacking variations in star formation that often occur. Regardless, the simplified prescription for metal-enriched star formation is often able to reproduce SFHs in the mass range of $10^{6}$-$10^{8}$ $M_{\mathrm{\odot}}$ remarkably well.

\subsection{Total Star Formation Rate} \label{subsec:total_SF}

Extending from the individual star formation histories discussed in the previous section, we also compare star formation rate densities averaged throughout the simulation box to evaluate how well 10 different realizations of our model reproduce star formation across entire \emph{Renaissance} volumes.

\begin{figure*}
\epsscale{1.15}
\plotone{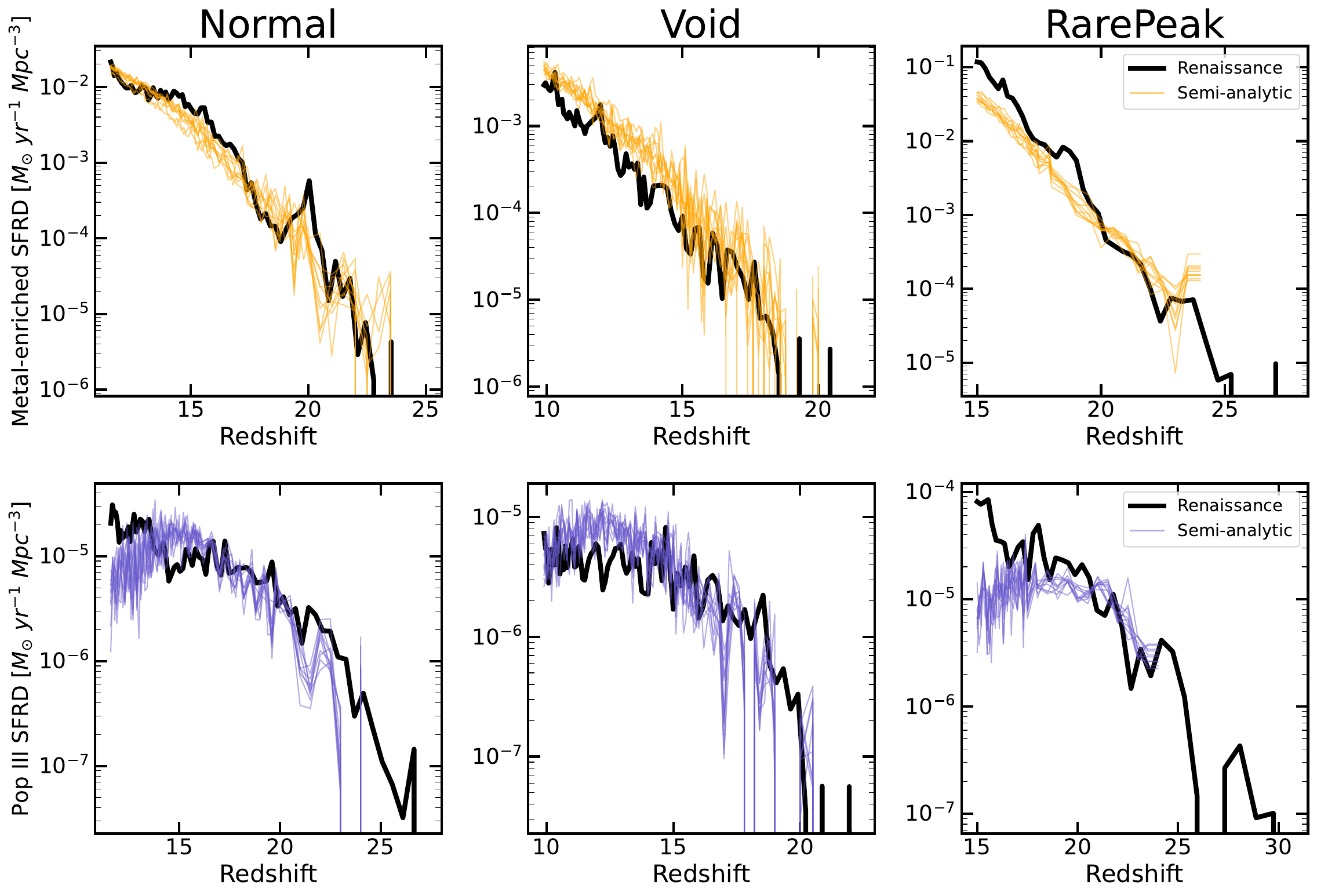}
\caption{Cosmic SFRDs for both metal-enriched (orange top-row) Population III (blue bottom-row) compared to \emph{Renaissance} (black). Results for the Normal, Void and RarePeak volumes are shown in the left-, middle-, and right-columns respectively. Overall, our model produces metal-enriched SFRDs typically within a factor of $\sim$2 of \emph{Renaissance} for a variety of environments. The larger discrepancy found at lower redshift in Void and RarePeak will be investigated in future work. \label{fig:SAM_SFRD}}
\end{figure*}

First, we will focus on metal-enriched star formation shown in the top row of Figure \ref{fig:SAM_SFRD}. In the top-left panel of Figure \ref{fig:SAM_SFRD}, we find that our model produces a metal-enriched SFRD consistent with the \emph{Renaissance} Normal region within a factor of $\sim$2. The good agreement with the \emph{Renaissance} SFRD indicates our two-stage metal-enriched star formation model, when aggregated for many halo SFHs, is able to closely reproduce total star formation in a large volume.

In addition to our simulations using the calibrations to the Normal region, we also consider the Void and RarePeak regions not included in our sample of halos (mentioned in Section \ref{sec:renaissance}). To ensure consistency with \emph{Renaissance}, we introduce a different LW background produced by Pop III stars used in \cite{2012wise_galaxy_birth_1} for the RarePeak region. For the Void region, we use the same LW background as the Normal region (see Section \ref{subsubsec:findings_popIII}).

Results for the Void region are shown in the middle column of Figure \ref{fig:SAM_SFRD}. The metal-enriched SFRD of the model is within a factor of $\sim$2 of \emph{Renaissance} for early times and at most differs by a factor of $\sim$3. For $z<16$, the steady stage begins in an increasing number of halos and will begin to dominate the metal-enriched SFRD. The scatter in the metal-enriched SFRD produced by our model begins to narrow as a consequence of the smooth SFHs produced during the steady stage mentioned in Section \ref{subsec:individual_SFHs}. This results in an SFRD that exhibits much less variation than \emph{Renaissance} and could explain the difference with the semi-analytic model at $z<16$.

In the right column of Figure \ref{fig:SAM_SFRD} we show our model results for the RarePeak region. Metal-enriched star formation in \emph{Renaissance} agrees within a factor of $\sim$2 for $z>19$ but eventually is more efficient than our model by a factor of $\sim$3 by $z \sim 15$. The source of this discrepancy is currently unknown, but we speculate a potential explanation is some physics missing from our prescription. For example, it has been shown that accretion driven mergers can boost star formation by increasing torques and driving gas towards the center of a halo \citep{2013hopkins}. This could explain the discrepancies we find in both Void and RarePeak.

Next, we will discuss Pop III star formation. Again, it is important to note that \emph{Renaissance} has difficulties in resolving halos with masses $\lesssim 7 \times 10^{6} M_{\mathrm{\odot}}$. Therefore, we focused mainly on calibration without going into the same level of detail as we do for metal-enriched star formation. As shown in the bottom row of Figure \ref{fig:SAM_SFRD}, the Pop III SFRDs produced by the semi-analytic model for Normal, Void, and RarePeak demonstrate similar behavior. The Pop III SFRD first increases, then, rests at the same value before gradually decreasing. This is due to small discrepancies in the $M_{\mathrm{min}}$ prescription in the semi-analytic model compared to \emph{Renaissance}.

We conclude that our model is able to reproduce the \emph{Renaissance} SFRDs within a factor of $\sim$2 for the Normal region. For the Void and RarePeak regions not used during the calibration of our model, the agreement is similar to the Normal at high-$z$ and varies at low-$z$ to a factor of $\sim$2 and $\sim$3 for Void and RarePeak, respectively. In future work, incorporating additional physics dependent on environmental effects could improve the agreement between our model and \emph{Renaissance}.

\subsection{Comparison with Updated Prescription for Population III Stars} \label{subsec:update_popIII}

Another question interesting to consider is the impact of incorporating recent findings updating the minimum halo mass of Population III star formation, $M_{\mathrm{min}}$, with new physics. In \cite{2021kulkarni}, $M_{\mathrm{min}}$ is obtained as a function of LW background (including $H_{\mathrm{2}}$ self-shielding), redshift, and baryon-dark matter streaming. While we were able to calibrate to the \emph{Renaissance} Pop III SFRD within a factor of $\sim$2, it is important to emphasize again that \emph{Renaissance} is limited in resolving minihalos that form Pop III.

\begin{figure}
\epsscale{1.15}
\plotone{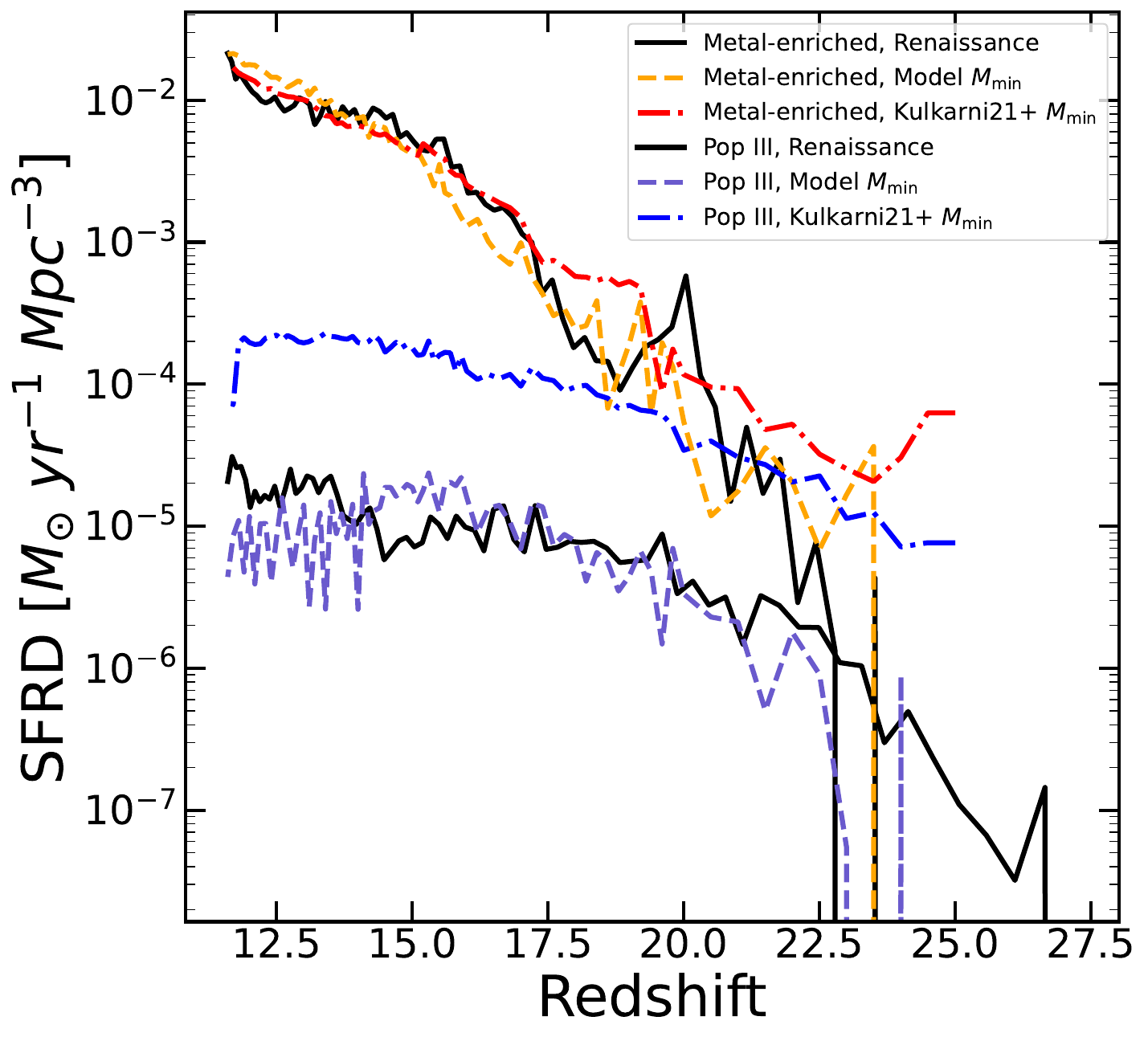}
\caption{Cosmic Metal-enriched and Population III SFRDs compared with \emph{Renaissance} for different $M_{\mathrm{min}}$ prescriptions in our model. The $M_{\mathrm{min}}$ with the mass resolution limit used in this paper (dashed) produces a Pop III SFRD that is roughly an order of magnitude lower than the Pop III SFRD using $M_{\mathrm{min}}$ from \cite{2021kulkarni}. The metal-enriched SFRDs agree within a factor of $\sim$2 except for $z>18$. \label{fig:kulkarni2021_Mmin}}
\end{figure}

When updating our prescription for $M_{\mathrm{min}}$ to be consistent with \cite{2021kulkarni} and removing the mass resolution limit, we find in Figure \ref{fig:kulkarni2021_Mmin} a Pop III SFRD approximately an order of magnitude higher than the prescription calibrated to \emph{Renaissance}. This is due to the inclusion of $H_{\mathrm{2}}$ self-shielding which lowers the impact of $J_{\mathrm{LW,21}}$ on $M_{\mathrm{min}}$. The impact on metal-enriched star formation due to the increased amount of Pop III was marginal. The metal-enriched SFRD from our semi-analytic prescription both with and without the updated $M_{\mathrm{min}}$ along \emph{Renaissance} agree within a factor of $\sim$2 for $z<18$. This result can be heavily influenced by any changes to LW feedback and reionization along with localized metal enrichment of halos not included in our model. Determining the relative impact of these processes after further calibration to \emph{Renaissance} will be the goal of future work. Our method will allow this work to be done with minimal computational expense and is a key benefit of semi-analytic models.

\subsection{H \textsc{ii} Regions} \label{subsec:HII_regions}

To determine the size and growth of ionized H \textsc{ii} regions we apply the same FFT method as \cite{2020visbal}. Using a $256^{3}$ cubic grid, we find the total number density of ionizing photons produced by dark matter halos that have escaped in the surrounding IGM in each cell. This is calculated as $P_{\mathrm{cell}} = f_{\mathrm{esc,II}} \eta_{\mathrm{II}} M_{\mathrm{\star,II,cell}} / (m_{\mathrm{proton}} V_{\mathrm{cell}}) + f_{\mathrm{esc,III}} \eta_{\mathrm{III}} M_{\mathrm{\star,III,cell}} / (m_{\mathrm{proton}} V_{\mathrm{cell}})$, where $f_{\mathrm{esc,II}}$ and $f_{\mathrm{esc,III}}$ are the escape fractions of hydrogen ionizing photons from halos in a cell containing metal-enriched and Pop III stars. The number of hydrogen ionizing photons produced per baryon for metal-enriched and Population III stars is assumed to be $\eta_{\mathrm{II}} = 4,000$ and $\eta_{\mathrm{III}} = 65,000$ respectively. The metal-enriched value corresponds to a Salpeter IMF from $0.1$ to $100M_{\mathrm{\odot}}$ and metallicity $Z=0.0004$ (see Table 1 in \cite{2007samui}). Note, if values were used for $\eta$ that correspond directly with \emph{Renaissance}, we would expect $f_{\mathrm{esc}}$ to change accordingly but the overall ionization fraction and structure would remain the same. The Pop III value is expected to emitted from a $\sim40M_{\mathrm{\odot}}$ star over its lifetime \citep{2002popIII_properties}. Both $M_{\mathrm{\star,II,cell}}$ and $M_{\mathrm{\star,III,cell}}$ are the total stellar masses ever formed by halos in a cell. Finally, $V_{\mathrm{cell}}$ is the comoving volume of a single cell. For each cell, $P_{\mathrm{cell}}$ is then smoothed over a range of spherical bubble sizes and is selected as the center of an H \textsc{ii} bubble if it exceeds a threshold of $i_{\mathrm{thres}} = 4 \bar{n}_{\mathrm{H}}$ \citep{2020visbal}.

\begin{figure}
\epsscale{0.98}
\plotone{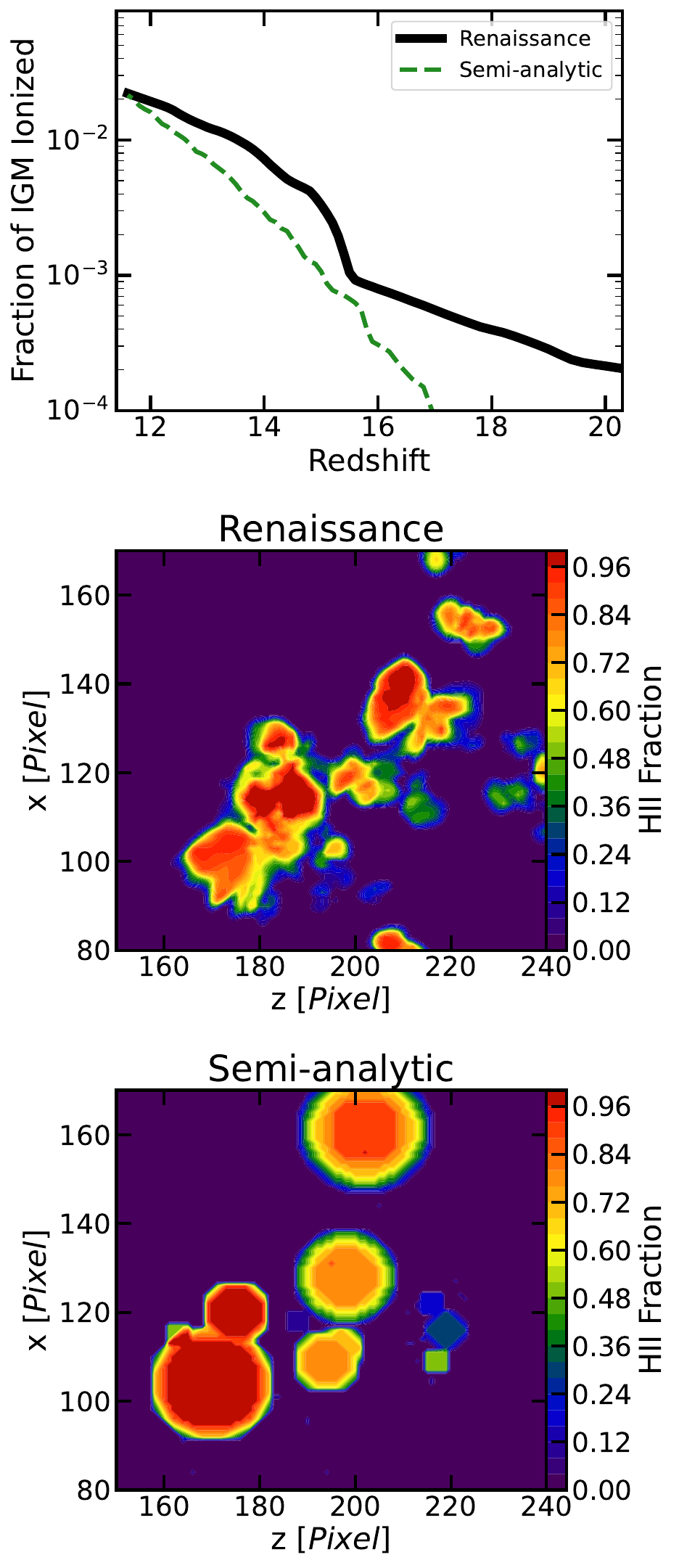}
\caption{Top panel: mean fraction of the volume of the IGM that is ionized in \emph{Renaissance} (black) compared with the the semi-analytic model (green). Projections of the H \textsc{ii} fraction for a region with a width and depth of $\approx$2.7 and $\approx$0.33 comoving Mpc, respectively. Middle panel: \emph{Renaissance}. Bottom panel: semi-analytic model. \label{fig:HII_regions}}
\end{figure}

The extent of ionized H \textsc{ii} regions determines when halos first form stars through the minimum halo mass for Pop III star formation and if a halo's descendants ever form metal-enriched stars in our model with no external enrichment. The top panel of Figure \ref{fig:HII_regions} compares the total ionization fraction in our model to \emph{Renaissance}. We calibrated the escape fraction of ionizing radiation to $f_{\mathrm{esc},\mathrm{II}} = 0.02$ for metal-enriched stars and $f_{\mathrm{esc},\mathrm{III}} = 0.1$ for Population III stars to produce a total fraction of the IGM ionized by our model within a factor of 2 of \emph{Renaissance} by $z \sim 11$.

In the middle and bottom panels of Figure \ref{fig:HII_regions}, we compare ionized bubble sizes in our model and \emph{Renaissance} for halos in the same thin volume at $z=11.6$. The difference in the H \textsc{ii} regions produced by our model in this region demonstrates the limitations of our current approach in reproducing ionization in simulations like \emph{Renaissance}. We find that our model typically produces larger ionized bubbles than \emph{Renaissance}. These H \textsc{ii} regions extend from massive halos with substantial star formation and are generally found at the center of clustered environments. To address this discrepancy, we are currently improving the prescription to produce more realistic bubbles and will be the focus of future work (Behling et al., in prep).

\section{Summary and Discussion} \label{sec:discussion}

In this paper we explored calibrating a semi-analytic model to the \emph{Renaissance Simulations}, radiative hydrodynamical simulations of the first stars and galaxies. An accurate semi-analytic model can make computationally efficient predictions in a fraction of the time, finishing in less than a day on a laptop compared to the $\sim$10 million core hours on the Blue Waters supercomputer used for a single \emph{Renaissance} simulation. The efficiency of semi-analytic models allows them to complement other simulations by extending over larger volumes, reaching lower redshifts, and exploring parameter space (e.g., Pop III IMF). While this work is focused on the \emph{Renaissance Simulations}, the same methodology can be straightforwardly implemented in other semi-analytic calculations and readily applied to other simulations.

To construct a semi-analytic model consistent with \emph{Renaissance}, we assembled a sample of star formation histories and calibrate our model to them. For metal-enriched star formation, we introduced a two-stage model (that does not include external enrichment) consisting of an early bursty formation stage that transitions to a steady stage with star forming at a constant efficiency. Notably, we implemented stochastic elements during star formation not typically included in other models by sampling parameters from the distributions in Table \ref{tab:overview_table}. We find that our stochastic two-stage model typically produced metal-enriched star formation histories consistent with \emph{Renaissance}, but completely capturing the variance in individual SFHs was not the intent of our model.

To calibrate Population III star formation, we implemented a minimum halo mass consistent with when the first stars form in \emph{Renaissance}. Once halos exceed this mass, a total Pop III stellar mass was assigned to halos along with a delay period before the onset of metal-enriched star formation. These values are sampled from the distributions described in Table \ref{tab:overview_table}. Our approach can be used to easily calibrate to different Pop III prescriptions and incorporate them into the model.

We determined how closely our model is able to reproduce the \emph{Renaissance} cosmic star formation history by comparing evolution of SFRDs and found agreement within a factor of $\sim$2 for redshifts $11 < z < 25$ for halo masses $\lesssim$$10^{9} M_{\mathrm{\odot}}$. We also tested our model on both the rarefied Void and overdense RarePeak regions in \emph{Renaissance}. The metal-enriched SFRDs produced by our model typically agrees by a factor of $\sim$2 in Void and $\sim$3 in RarePeak, respectively. The difference in SFRDs based on environment suggests there is physics not currently included in our model that either leads to higher star formation efficiencies or increased external enrichment. Therefore, we expect semi-analytic models that do not consider these effects will fail to reproduce star formation by similar factors.

Additionally, while reproducing the \emph{Renaissance} star formation was a primary goal, our Population III prescription could be updated to include more recent findings. For example, it has been shown that baryon-dark matter streaming, self-shielding from LW radiation, and X-ray feedback impact the $\mathrm{H_{2}}$ fraction and therefore halo minimum mass for Pop III formation. A combination of these effects can raise the minimum halo mass and delay the formation of the first stars. We found using the minimum dark matter halo mass for Pop III formation from \cite{2021kulkarni} that the Pop III SFRD increases by about an order of magnitude compared to \emph{Renaissance}. Meanwhile, for $z<18$ the metal-enriched SFRDs agree within a factor of $\sim$2. Both LW feedback and metal enrichment could impact star formation and will be the focus of future work implementing additional calibrations to \emph{Renaissance}.

Using the grid-based approach of \cite{2020visbal} to calculate feedback from LW radiation using fast Fourier transforms, we found our model produces total ionized fractions that agree within a factor of $\sim$3 to \emph{Renaissance} between redshifts $14 \lesssim z \lesssim 11$ and even closer agreement between $18 \lesssim z \lesssim 14$. When comparing individual H \textsc{ii} regions, we found that our model typically forms larger ionized bubbles than those in \emph{Renaissance}. H \textsc{ii} regions in highly clustered environments can suppress the onset of star formation in neighboring halos and could reduce the overall amount of metal-enriched and Population III stars.

In future work, we plan to investigate the impact of the environment on star formation in greater detail. Including dependence on large scale overdensity, external enrichment, and a more detailed treatment of ionized H \textsc{ii} bubble formation in our calibration could potentially improve the agreement with \emph{Renaissance} for different environments. Our two-stage prescription for metal-enriched star formation would then be more accurate for a range of cosmological environments. 

We also intend to make observational predictions by applying our calibrated model to dark matter merger trees that extend over a large statistically representative volume. Some of these predictions include supernovae rates \citep[e.g.,][]{2005scannapieco,2005weinmann,2005wise,2006mesinger,2012pan,2012hummel,2013johnson,2013tanaka,2013whalen_2,2014deSouza,2018hartwig}, luminosity functions \citep[e.g.,][]{2023bouwens,2023atek}, and abundance predictions in strong gravitational lenses. Incorporating direct collapse black hole seeds in our model and tracking their growth could also potentially lead to new insights into the supermassive black holes found at the centers of most galaxies.

\begin{acknowledgments}

\section{Acknowledgments} \label{sec:acknowledgments}
We thank Greg Bryan and Zoltan Haiman for useful discussions. R.H. and E.V. acknowledge support from NASA ATP grant 80NNSSC22K0629 and NSF grant AST-2009309. J.W. acknowledges support by NSF grant AST-2108020 and NASA grants 80NSSC20K0520 and 80NSSC21K1053. The numerical simulations in this paper were run on the Ohio Supercomputer Center (OSC).

\end{acknowledgments}

%

\vspace{5mm}
\facilities{Ohio Supercomputer Center (OSC)}


\software{\textsc{numpy} \citep{2020NumPy}, 
          \textsc{scipy} \citep{2020SciPy},
          \textsc{matplotlib} \citep{2007hunter},
          \textsc{cython} \citep{2011Cython}, 
          \textsc{jupyter notebook} \citep{2016Jupyter}, 
          \textsc{rockstar} \citep{2013behroozi_a}, 
          \textsc{consistent trees} \citep{2013behroozi_b}, 
          \textsc{yt} \citep{2011yt}, 
          \textsc{ytree} \citep{2019ytree}.
          }




\bibliography{sample631}{}
\bibliographystyle{aasjournal}



\end{document}